\documentclass[%
 aip,
 amsmath,amssymb,
preprint,%
]{revtex4-1}

\usepackage{graphicx}
\usepackage{dcolumn}
\usepackage{bm}
\usepackage[utf8]{inputenc}
\usepackage[T1]{fontenc}
\usepackage{mathptmx}
\usepackage{etoolbox}
\usepackage{xcolor}
\usepackage{soul}

\makeatletter
\def\@email#1#2{%
 \endgroup
 \patchcmd{\titleblock@produce}
  {\frontmatter@RRAPformat}
  {\frontmatter@RRAPformat{\produce@RRAP{*#1\href{mailto:#2}{#2}}}\frontmatter@RRAPformat}
  {}{}
}%
\makeatother
\begin{document}

\title{Energetics of the Charge Generation in Organic Donor-Acceptor Interfaces}
\author{Artur M. Andermann}
 \affiliation{Department of Physics, Universidade Federal de Santa Catarina, 88040-900, Florianópolis, Santa Catarina, Brazil}
\email{artur.andermann@posgrad.ufsc.br}
\author{Luis G. C. Rego}
 \email{luis.guilherme@ufsc.br}
\affiliation{Department of Physics, Universidade Federal de Santa Catarina, 88040-900, Florianópolis, Santa Catarina, Brazil}


\begin{abstract}
Non-fullerene acceptor (NFA) materials have posed new paradigms for the design of organic solar cells (OSC), whereby efficient carrier generation is obtained
with small driving forces, in order to maximize the open-circuit voltage (V$_{\text{OC}}$).
In this paper we use a coarse-grained mixed quantum-classical method, that combines Ehrenfest and Redfield theories, to shed light on 
charge generation process in small energy offset interfaces.
We have investigated the influence of the energetic driving force as well as the vibronic effects on the charge generation and photovoltaic energy conversion.  
By analyzing the effects of the Holstein and Peierls vibrational couplings, we find that vibrational couplings produce an overall effect of improving the charge generation. 
However, the two vibronic mechanisms play different roles: 
the Holstein relaxation mechanism decreases the charge generation whereas the Peierls mechanism always assists the charge generation. 
Moreover, by examining the electron-hole binding energy as a function of time, we evince two distinct regimes for the charge separation: the temperature independent 
excitonic spread on a sub-100 fs timescale and the complete dissociation of the charge-transfer state that occurs on the timescale of tens to hundreds of picoseconds, 
depending on the temperature.
The quantum dynamics of the system exhibits the three regimes of the Marcus electron transfer kinetics as the energy offset of the interface is varied.
\end{abstract}

\maketitle


\section{Introduction}

The photovoltaic conversion  process in organic solar cells (OSC) consists of several steps that affect the overall efficiency of
the device.\cite{bredas2009molecular, Durrant2010, gao2020critical, jiang2019understanding, Armin2021, yavuz2017dichotomy} 
Charge photogeneration has always been a major concern regarding cell efficiency,\cite{Durrant2010} 
so that energy cascades were considered essential for overcoming the Coulomb attraction and generating free carriers as quickly as possible
to prevent losses from recombination.
Consistent progress has been achieved by optimizing the transport properties of the photoactive materials as well as  the interface morphology between 
them.\cite{Durrant2010,Armin,Armin2021}
The first efficient solar cells were fabricated with fullerene derivatives as acceptor materials, as they provide
remarkable electron-accepting and electron-transporting capabilities, low reorganization energy and good miscibility.\cite{Durrant2010,Hou2018}
However, improvements in cell efficiency of fullerene-based OSC saturated around 12\%, whereas 
the photovoltaic efficiency of non-fullerene OSC has increased strongly, nearing the 20\% 
efficiency mark.\cite{Hou2017,Armin2021}
The photoconversion efficiencies of fullerene OSC saturated mostly due to excessive energy losses undergone at the donor-acceptor (D-A) interface region. 
Since the exciton binding energy is stronger in organic molecules than in inorganic crystalline materials, the energetic driving force of the D-A interface was
considered essential for generating the charge separated state.\cite{hallermann2008charge,menke2014exciton} 
Moreover, guidelines based on fullerene-based OSC indicated that the photocurrent should fall abruptly for small charge separation 
driving forces.\cite{li2015high} 
Such a notion was supported by the analogous charge separation process in natural photosynthesis, \cite{mimicking} in which case a strong energetic driving 
force is required to guarantee free charge generation.
Natural photosynthesis, however, has not evolved for producing energy at high efficiencies.\cite{Blankenship}

The use of non-fullerene acceptors (NFA) has made it possible to increase the photovoltaic efficiency of organic solar cells to levels comparable to the inorganic
counterparts. This fact reveals the importance of the basic molecular processes to the efficiency of organic photovoltaics, 
in addition to morphological and macroscopic design rules of the device.
The improvements have been achieved empirically, though, without a complete understanding of the underlying fundamental mechanisms.
{\cite{menke2018understanding,Qian2018,Nakano2019}}
It is intriguing, for instance, how a small energetic driving force gives rise to efficient charge photogeneration without the need of
external electric fields.{\cite{Andrienko}}
Motivated by that,
recent studies have proposed models to explain the reduced non-radiative voltage losses in systems with low D-A energetic 
offsets.{\cite{Benduhn2017,Qian2018,JennyNelsonJACS,Chen2021}}
Another open issue is  the influence of entropic as opposed to enthalpic effects on the charge separation as the driving force continues to 
decrease.{\cite{Gelinas,Grancini,Kaake,Mukamel}}
In that respect vibronic couplings should play a significant role.{\cite{Chen2018}}
Experimental evidence has shown the relevance of vibronic effects during the charge separation and charge transport  processes.
{\cite{falke2014coherent, song2014vibrational, van2020vibronic, de2017vibronic, kilina2015light, de2016tracking}} 
In essence, the understanding of the relevant underlying  mechanisms for organic photovoltaics can open new paths in the development of material blends with better 
performance.{\cite{de2017vibronic}}

In this work  we address some of these issues by means of theoretical simulations of the charge generation dynamics in a 
model D-A interface, 
taking into account quantum mechanical and dissipative effects.
The  process  is modelled on the basis of the system-bath partition approach. The degrees of freedom directly associated with the process of photoinduced charge
generation, namely the electron and hole quantum states and their vibrational reorganization modes, are described coherently
within the framework of the Ehrenfest method. The system-bath coupling is described by the Redfield theory for dissipative quantum dynamics.
We have studied the influence of the energetic driving force and vibronic effects on the charge generation and photovoltaic energy conversion.  
By analyzing the effects of the Holstein and Peierls vibrational couplings, we find that vibrational couplings produce an overall effect of improving the charge generation. 
However, the two vibronic mechanisms play different roles: 
the Holstein relaxation mechanism decreases the charge generation whereas the Peierls mechanism always assists the charge generation. 
We examine the electron-hole binding energy as a function of time and evince two distinct regimes for charge dissociation: the temperature independent 
excitonic spread on the sub-100 fs timescale and the charge-transfer state dissociation on the timescale of tens of picoseconds, so that when the electron-hole
pair reaches the interface its binding energy is much smaller that the initial excitonic binding energy.
Last, we investigate the effect of a macroscopic electric field on the charge dissociation process, as a function of the D-A energy offset.

\section{Theory and Methods}

By adopting the system-bath approach we partition the hamiltonian of the charge generation process  as 
\begin{equation}
H = H_{S} + H_{B} + H_{SB},
\end{equation} 
where the system hamiltonian, $H_S$, comprises the degrees of freedom directly associated with the process of photoinduced charge generation, 
namely the electron and hole quantum states and their vibrational reorganization modes, herein described as classical degrees of freedom. 
Therefore, $H_S$ is actually a mixed quantum-classical hamiltonian treated within the 
framework of the Ehrenfest method. The environment hamiltonian, $H_B$, accounts for all the remaining degrees of freedom of the D-A interface and its environment, 
namely the vibrational degrees of freedom and the fluctuations of the dielectric background, which we describe as an ensemble of quantum harmonic oscillators. 
Finally, $H_{SB}$ accounts for the interactions between the electron-hole pair and the environment.
The system-bath coupling is described within the framework of the Redfield theory. 
The stationary Redfield formalism is based on the Markovian approximation and, therefore, it does not describe coherent vibronic couplings. The Ehrenfest method, on the other hand, describes the coherent electron-phonon interactions at the semiclassical level. We use both electron coupling approaches to complement each other.

The combination of different methodologies is generally adopted to treat the dynamics of complex systems. 
For example, the Marcus-Jortner-Levich (MJL) theory{\cite{Jortner,MJL}} describes the environment in terms of intramolecular vibrations (inner-sphere) and 
solvent/dielectric fluctuations (outer-sphere); in addition, a cutoff frequency can be set to separate classical (low energy) from quantum 
(high energy) bath/environment modes. Another example, more closely related to the present method, is the Reduced Density Matrix Hybrid (RDMH) 
approach, {\cite{RDMH-I,RDMH-II}} 
according to which the system's environment is partitioned into core modes and reservoir modes. The core modes are treated quantum mechanically at the level of 
a perturbative quantum master equation whereas the reservoir modes are described within the framework of the Ehrenfest method, in order to account for the low 
energy (slow) modes of the environment. The Ehrenfest dynamics is known for describing coherent non-Markovian effects produced by non-equilibrium vibrations. 
Thus, the proposed combination of methods (quantum dissipative with semiclassical Ehrenfest) is well suited for describing the superposition of coherent effects 
and dissipative dynamics. Our method aims to accomplish this goal for the excited-state dynamics of molecular crystals, as opposed to electronic transfer in 
solution chemistry.
Our analysis aims at investigating the primary events that occur after the photoexcitation of the electron-hole pair in the vicinity of the D-A interface, 
as due to the intrinsic energetic properties of the heterojunction, it does not provide a description of the net current flow in the device.
Next we describe the model system for the D-A interface and,
then, the implementation of the Redfield equations.

\subsection{Donor-Acceptor System}

\begin{figure}[h!]
    \centering
  \includegraphics[width = 9.5cm]{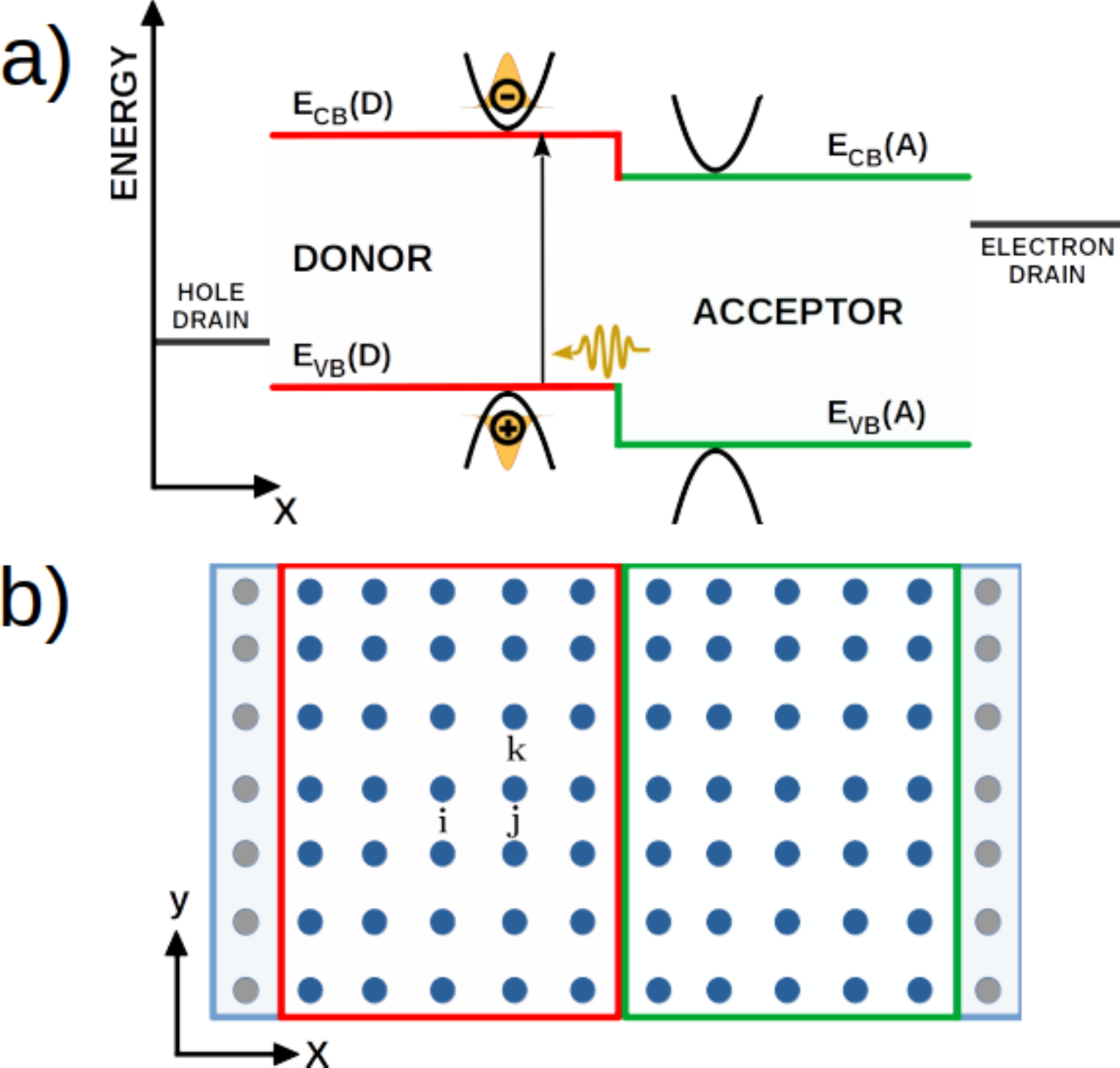}
  \caption{Model representation of the donor-acceptor interface. Panel a) depicts the energetics of the D-A interface. The outermost layers are the electron and hole
drains. The vertical arrow indicates the photoexcitation of the electron-hole pair. The paraboloid describes the confining potential energy
surface of the molecular site. 
Panel b) depicts the spatial configuration of the D-A interface. Each molecule is described as a site in a 2D 
lattice. Electron and hole can coexist on the same site, or occupy separate sites.} 
 \label{DAmodel} 
\end{figure}

As shown in Figure \ref{DAmodel}-a), the D-A interface is modelled as a two-dimensional (2D) lattice, with
energy profile corresponding to a staggered (type II) interface.
Upon photoexcitation of a molecular site the electron-hole pair diffuses to the interface, where it gives rise to a charge transfer (CT) state. 
Herein, we assume the photoexcitation of a molecular site in the donor material, however, since the model has particle-hole symmetry, the  forthcoming 
results are equally valid for an exciton created in the acceptor material.
If the CT state overcomes the electron-hole binding and dissociates before recombination -- either radiative or non-radiative -- 
annihilates the CT state, a pair of 
free charge carriers with energy $E_{CS}=E_{CB}(A)-E_{VB}(D)$ is produced, with $E_{CS}$ designating the energy of the 
charge separated state. 
The energy difference $E_g^{\text{opt}} - E_{CS}$ can be associated with the energetic driving force of the charge generation process.\cite{Nakano2019,menke2018understanding}
In the present model, the electron and hole populations are eventually collected at the respective drain layers, here positioned  
0.3 eV below $E_{CB}(A)$ and 0.3 eV above $E_{VB}(D)$, in analogy to the cathode and anode terminals. 
The spatial arrangement of the  2D  D-A interface model is shown in Figure \ref{DAmodel}-b), where each site of the lattice represents a molecular site. 
Periodic boundary conditions are applied along the $y$ direction. 

\begin{figure}[htbp]
    \centering
  \includegraphics[width = 0.55\linewidth]{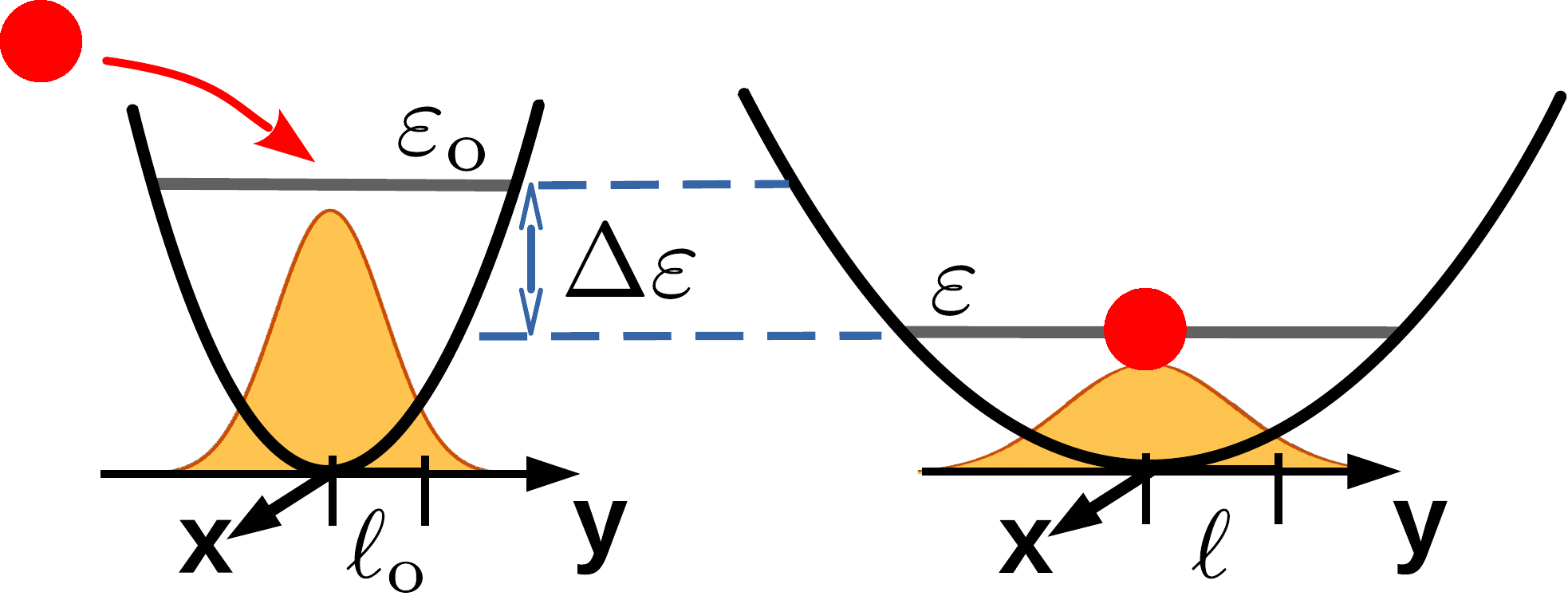}
  \caption{Schematics of the electronic relaxation. The parabola describes the confinement potential felt by the electron (or hole) with
its gaussian wavefunction profile. 
On the left-hand  side we  show an empty molecular site, with confinement energy $\varepsilon_\text{o}$ and the corresponding confinement length $\ell_\text{o}$.
On the right-hand  side we  show an occupied molecular site, with on-site energy  $\varepsilon$ and $\ell$. 
The electronic relaxation energy is $\Delta \varepsilon=\varepsilon_\text{o}-\varepsilon$.  }
 \label{reorganization} 
\end{figure}

The paraboloids depicted in Figure \ref{DAmodel}-a) describe the local confinement potential felt by the electron 
and the hole at a given molecular site. 
We assume that the parabolic confinement potential undergoes a vibrational reorganization whenever charge is transferred in or out of the molecular site 
(see Figure \ref{reorganization}). This mechanism gives rise to coherent vibronic effects that we describe within the framework of the Ehrenfest method.
The density profile of the electron and hole in the molecular site is given by the gaussian
\begin{equation}
g(\vec{r}-\vec{R}) = 
\sqrt{\frac{2}{\pi \ell^2}}\exp\left[-\Big(\frac{\vec{r}-\vec{R}}{\ell}\Big)^2 \right], 
\end{equation}
with $\vec{R}$ designating the position of the molecular site, and  $\ell$ is the confinement radius of the molecular site, given by 
\begin{equation}
\ell = \sqrt{2\hbar^2/(m_e\varepsilon)} 
\label{confine}
\end{equation}
for a gaussian wavepacket of on-site energy $\varepsilon$. 

To write the system hamiltonian, $H_S$, we assume an orthogonal basis set comprised of diabatic electronic states $\{|i\rangle\}$ associated with the molecular sites:
\begin{equation}
H_S \equiv H^{\text{el/hl}} = \sum_{i}^{N}	\left\{E_i^{\text{el/hl}} + \varepsilon_i(t) - \Phi^{\text{el/hl}}_i(t)\right\} |i \rangle \langle i |  
+ \sum_{i \neq j}^{N} V_{ij}(t) |i \rangle \langle j |,
\label{Hs}
\end{equation} 
with $N$ denoting the total number of molecular sites in the lattice. 
Hereafter, we use the symbols \{$|i\rangle,|j\rangle\}$ to designate exclusively the local diabatic quantum states.
The on-site terms of $H_S$ are defined as
 $E^{\textbf{el}}_i = E^{CB}_i$, $E^{\textbf{hl}}_i = E^{VB}_i$, 
$\varepsilon_i$ is the confinement energy associated with site $i$, as given by Eq. (\ref{confine}), 
and $\xi^{\text{el/hl}}_i$ represents the electron-hole electrostatic interaction, given by the mean-field potential 
\begin{equation}
\Phi^{\text{el/hl}}_i(t)  = \xi_{bind}~ \left(\sum_j \frac{P^{\text{hl/el}}_j}{(1 + d_{ij})}\right).
\label{bind}
\end{equation}
In Eq. (\ref{bind}), $\xi_{bind}$ is assumed to be the same for the electron and hole excitations.  $P^{\text{el}}_j$ ($P^{\text{hl}}_j$) designates 
the average electronic (hole) population on site $j$ and $d_{ij}=|\vec{R}_i-\vec{R}_j|$ is the distance between sites $i$ 
and $j$.  Thus, the total electron-hole binding energy is given by $E_{\text{bind}} = \text{Tr}[\sigma^\text{el} \Phi^\text{el}  + \sigma^\text{hl} \Phi^\text{hl}]$, 
with $\sigma$ representing the reduced density matrix.

The coupling $V_{ij}$ between lattice sites  is given  by 
\begin{equation} 
V_{ij}  = V_0~F(\ell_{i},\ell_{j},d_{ij}) =  
V_0 \left[
 \frac{ 2 \ell_{i}\ell_{j} }{ \ell_{i}^2+\ell_{j}^2} \exp \Big ({ \frac{- d^2_{ij}}{\ell_{i}^2+\ell_{j}^2} } \Big ) 
\right],
\label{hopping}
\end{equation}
where $V_0$ is the bare electronic coupling and the form factor $F$ results from the overlap between
gaussian wavepackets located at lattice sites $i$ and $j$
\begin{equation}
F(\ell_i,\ell_j,d_{ij}) = \int g_i(\vec{r}-\vec{R}_i)~g_j(\vec{r}-\vec{R}_j) dxdy.
\end{equation}
The time dependence of $\varepsilon_i(t)$ and $V_{ij}(t)$, which comprise Eq. (\ref{Hs}),  gives rise to {\it intra}-molecular and {\it inter}-molecular 
vibronic couplings that we associate with the Holstein and Peierls couplings, respectively.

We describe the coherent coupling between the electron-hole states and the intra-molecular vibrational degree of freedom via the 
Ehrenfest method. The vibrational hamiltonian coupled to the electron-hole in the site $i$ is given by
\begin{equation}
h_i = \frac{p_i^2}{2\mu} + \frac{\mu \Omega^2}{2}(\ell_i-\ell_{\text{o}})^2 + W^{eh} + Q^{thermo},
\label{intra}
\end{equation}
where $\ell_i$ is the vibrational coordinate associated with the molecular site $i$  and $p_i$ is the conjugate momentum. In addition,
$\ell_{\text{o}}$ is the confinement length for an empty site, 
$\mu$ is the effective mass of the vibrational mode and $\Omega$ is the frequency of the relevant normal mode. The energy $W^{eh}$ is the work exchanged with the 
electronic degrees of freedom via the Ehrenfest force $F^{eh}$ and 
$Q^{thermo}$ is the heat exchanged with the classical thermostat. 
A similar hamiltonian  has been considered by Egorova et al.\cite{egorova2001modeling,egorova2003modeling}  for quantum intra-molecular degrees of freedom.
The Ehrenfest force that acts on the vibrational coordinate of a given molecular site $k$ is 
\begin{equation}
F^{eh}_k = F_k^{\text{el}} + F_{k}^{\text{hl}} =  - \frac{\partial U^{\text{el}}}{\partial \ell_k} - \frac{\partial U^{\text{hl}}}{\partial \ell_k}, 
\label{Ehrenfest1}
\end{equation}
with $U^{\text{el}} = Tr[\sigma^{\text{el}}H^{el}]$ and $U^{\text{hl}} = Tr[\sigma^{\text{hl}}H^{hl}]$.
Solving equation (\ref{Ehrenfest1}) we obtain (see Appendix A)
\begin{equation}
F^{eh}_k = - (\sigma^{\text{el}}_{kk} + \sigma^{\text{hl}}_{kk}  ) \frac{d \varepsilon_k}{d \ell_k} 
- \sum_{i \neq j} (\sigma^{\text{el}}_{ij} + \sigma^{\text{hl}}_{ij}) \frac{\partial V_{ij}}{\partial \ell_k}.
\label{Ehrenfest2}
\end{equation}
The first term on the right-hand side (RHS) of Equation (\ref{Ehrenfest2}) is responsible for the relaxation of the molecular site due to its charge occupation; 
this is the classical work exchanged between electronic and vibrational degrees of freedom. 
The second term gives rise to coherent vibronic effects, for it is proportional to
the off-diagonal elements of $\sigma$ and couples the electronic and nuclear coordinates via $\partial V_{ij}/\partial \ell_k$.\cite{Kewin}

The classical equations of motion derived from Eq. (\ref{intra}) are solved with the velocity Verlet algorithm, 
coupled to a classical bath, described by the Berendsen thermostat,\cite{berendsen1984molecular} with relaxation constant  $\gamma^{-1} = 0.1$ ps. 
The perturbing forces change the equilibrium configuration of the parabolic confining potential, which reacts with a restoring force
$F^\text{res} = -\mu\Omega^2(\ell - \ell_{\text{o}})$. 

\subsection{Redfield Coupling with the Environment}

A rigorous method to derive the equations of motion for the system hamiltonian $H_S$ coupled with the environment is based on the application of projection 
operators $\mathcal{P}$ and $\mathcal{Q}$, where $\mathcal{Q}\equiv1-\mathcal{P}$, and the Nakajima-Zwanzing equations.\cite{nakajima1958quantum,zwanzig1960ensemble,may2008charge}
The projection operator $\mathcal{P}$ is defined as $ \mathcal{P} \rho = \text{tr}_B(\rho)\otimes\rho_B=\sigma\otimes\rho_B$, 
where $\rho_B$ is the density operator of the bath
\begin{equation}
\rho_B = \frac{e^{-H_B/k T}}{\text{tr}_B \{  e^{-H_B / k T }  \} },
\end{equation} 
which is assumed to be in thermal equilibrium, that is $\text{tr}_B\{H_{SB}\rho_B\}=0$, and
\begin{equation}
H_B = \sum_{\alpha} \hbar \omega_\alpha ( b^\dagger_\alpha b_\alpha + \frac{1}{2} ).
\end{equation}
Here, $b_\alpha$ and $b^\dagger_\alpha$ designate the bosonic operators of the bath and $\omega_\alpha$ is the corresponding frequency. 
It is also assumed that system and bath are not correlated at the outset, $\rho(0)=\sigma(0)\otimes\rho_B(0)=\sigma(0)\rho_B(0)$.

Although the Nakajima-Zwanzing equations provide an exact framework for the dynamics of $H_S$, they are not computationally practical.
Thus, approximation steps are required to obtain the stationary Redfield equations used in this 
work.\cite{redfield1965theory,may2008charge,egorova2001modeling,egorova2003modeling}
These approximation steps consist of: 
(i) first, the system-bath coupling is treated perturbatively using the second order Born approximation; 
(ii) since the resulting equation is not local in time, the local approximation $\sigma(t')\rho_B \approx \sigma(t)\rho_B$  is invoked; 
and (iii) the formalism assumes that the dynamics of the system does not alter the bath significantly, which is considered to be in equilibrium. 
Then, the upper limit of the time integration is replaced by infinity, and a local-time equation of motion for the reduced density matrix.
All the above approximation steps are incorporated in Eq. (\ref{Red1})
\begin{equation}
\frac{\partial \sigma (t)}{\partial t } = 
-\frac{1}{\hbar^2} \int_{0}^{\infty} dt'~\text{tr}_B \Big \{[H_{SB}(t),[H_{SB}(t'),\sigma(t) \rho_B]] \Big \},
\label{Red1}
\end{equation}
where we have written $H_{SB}(t) = e^{\frac{i}{\hbar} (H_S + H_B) t} H_{SB} e^{-\frac{i}{\hbar} (H_S + H_B) t}$ in the interaction picture. 
Using the adiabatic representation for the system hamiltonian, $H_S |\varphi_a\rangle = E_a |\varphi_a\rangle$, the Redfield equation for the reduced density matrix $\sigma$ reads
\begin{equation}\label{EQT1}
\frac{\partial \sigma_{ab}(t)}{\partial t} = - i \omega_{ab} \sigma_{ab} + \sum_{c, d} R_{abcd} \sigma_{cd}(t)
\end{equation}
where $\omega_{ab}=(E_a-E_b)/\hbar$ designates the eigenfrequencies and $R_{abcd}$ is the Redfield relaxation tensor.\cite{redfield1957theory, redfield1965theory} 
The first term on the RHS accounts for the coherent quantum dynamics of S whereas the second term describes the interaction of 
S with the environment. The Redfield tensor can be written in terms of correlation functions of the system-bath coupling
\begin{equation}
R_{abcd} = \Gamma^+_{dbac} + \Gamma^-_{dbac} - \sum_{n} \Gamma^+_{annc} \delta_{db} - \sum_{n} \Gamma^-_{dnnb} \delta_{ac} 
\end{equation} 
where
\begin{eqnarray}
\Gamma^+_{abcd} = \frac{1}{\hbar^2} \int_{0}^{\infty} d \tau e^{- i \omega_{cd} \tau} 
\langle \langle a | \tilde{H}_{SB}(\tau) | b \rangle \langle c | H_{SB} | d \rangle \rangle_B \label{GammaMais}, \\ 
\nonumber \\
\Gamma^-_{abcd} = \frac{1}{\hbar^2} \int_{0}^{\infty} d \tau e^{- i \omega_{ab} \tau} 
\langle \langle a | H_{SB} | b \rangle \langle c | \tilde{H}_{SB}(\tau) | d \rangle  \rangle_B \label{GammaMenos}, 
\end{eqnarray}
with $\tilde{H}_{SB}(t)= e^{(i/\hbar)H_Bt} H_{SB} e^{-(i/\hbar)H_Bt}$ and 
$\langle \cdots \rangle_B$  designates the thermal average over the bath degrees of freedom. 

The coupling of electrons and holes with the quantum bath is described by the Holstein model,\cite{holstein1959studies1,  holstein1959studies2, mozafari2013polaron, coropceanu2007charge, lee2015charge, tempelaar2018vibronic, hsu2020reorganization} 
which has been used, together with the Peierls model, to describe polaronic effects in molecular crystals.\cite{Holstein-1959,Holstein-Polaron}
For the intra-molecular electron-phonon relaxation, we have
\begin{equation}
H_{SB} = \sum_{\alpha} \sum_{i} g_{i, \alpha} \hbar \omega_\alpha |i\rangle \langle i | (b^\dagger_\alpha + b_\alpha) 
\end{equation}
which contains the dimensionless parameter  $g_{i, \alpha}$ for electron-phonon coupling that is associated with the molecular site $i$ and the $\alpha$th bath mode. 
We assume that all sites have the same spectral density.
The strength of the system-bath coupling is given by $C'_{i, \alpha} \equiv g_{i, \alpha} \hbar \omega_\alpha $. 
Thus, from  Eqs.  \eqref{GammaMais} and \eqref{GammaMenos} we can derive the Redfield relaxation tensors
\begin{eqnarray}
\Gamma^+_{abcd} &=&  \sum_{i, j} \frac{\Lambda^i_{ab} \Lambda^j_{cd}}{\pi \hbar} \int_{0}^{\infty} d\omega J_{i j}(\omega) \int_{0}^{\infty} d \tau \Big [ \big (n(\omega) + 1  \big ) e^{-i ( \omega + \omega_{cd} ) \tau}  + n(\omega) e^{i (\omega - \omega_{cd}) \tau} \Big ] \label{CorrFunc6} , \label{gama1}\\
\Gamma^-_{abcd} &=&  \sum_{i, j} \frac{\Lambda^i_{ab} \Lambda^j_{cd}}{\pi \hbar}  \int_{0}^{\infty} d\omega J_{i j}(\omega) \int_{0}^{\infty} d \tau \Big [ \big (n(\omega) + 1  \big ) e^{i ( \omega - \omega_{ab} ) \tau}  + n(\omega) e^{-i (\omega + \omega_{ab}) \tau} \Big ] \label{CorrFunc7}. \label{gama2}
\end{eqnarray}
The expression for the $\Gamma$ rates, above, make use of the system-bath interaction spectral density (SD) 
$J_{ij}(\omega) = \pi \sum_{\alpha} \frac{C'_{i, \alpha}C'_{j, \alpha}}{\hbar} \delta(\omega - \omega_\alpha)$ (to be defined ahead) and the 
Planck thermal distribution $n(\omega) = \frac{1}{e^{\beta \hbar \omega} - 1 }$, calculated with the instantaneous subsystem S temperature $T$ and frequency $\omega$.
We also define $\sum_i \Lambda^i_{ab} \equiv \sum_i \langle \varphi_a | i \rangle \langle i | \varphi_b \rangle$, recalling that \{$|i\rangle,|j\rangle$\} designate the
diabatic on-site quantum states whereas the \{|$\varphi\rangle$\} states comprise the adiabatic basis set. 
 
To calculate Eqs. (\ref{gama1}) and (\ref{gama2}), we assume that the interaction spectral density is described by Drude-Lorentz distribution, $J(\omega) = 2 \lambda \frac{\omega \omega_c}{\omega^2 + \omega^2_c}$, and also that the phonon modes of different sites are uncorrelated
$J_{ij}(\omega) = J(\omega) \delta_{ij} $.\cite{Ishizaki, lee2015charge}
The Drude cutoff frequency was set to $\omega_c^{-1} = 25$ fs (equivalent to 1334 cm$^{-1}$, or $~0.165$ meV), due to the coupling of the electronic degrees of 
freedom with the high energy phonon modes 
associated to the stretch normal mode of the C$=$C bond.\cite{lee2015charge, hsu2020reorganization}
It determines the peak of the spectral density; when the relevant frequencies 
of the system are much lower than then $\omega_c$ the reservoir behaves like an Ohmic heat bath.
Considering these definitions, the reorganization energy can be related to $J(\omega)$ as  
\begin{equation}
\lambda = \frac{1}{\pi} \int_0^\infty d \omega \frac{J(\omega)}{\omega}.
\end{equation}

Using the one sided delta function 
definition $\int_{0}^{\infty} d\tau e^{\pm i ( \omega \mp \Omega) \tau} = \pi \delta(\omega \mp \Omega ) \pm i \mathcal{P} \frac{1}{\omega \mp \Omega}$, where $\mathcal{P}$ denotes the principal Cauchy value, the real part of Redfield tensor elements are given by
\begin{eqnarray} \label{GammaMaisDef}
\text{Re}[\Gamma^+_{abcd}] = \frac{1}{\hbar}\sum_i \Lambda^i_{ab} \Lambda^i_{cd} \begin{cases}
J(\omega_{cd}) n(\omega_{cd})  \ \ & \text{if} \ \ \omega_{cd} > 0 \\ 
J(\omega_{dc}) \big [ n(\omega_{dc}) + 1 \big ]   \ \ & \text{if} \ \ \omega_{dc} > 0 \\ 
\lim\limits_{\omega \rightarrow 0} J(\omega) n(\omega) \ \ & \text{if} \ \ \omega_{cd} = 0 
\end{cases} 
\end{eqnarray} 
\begin{eqnarray} \label{GammaMenosDef}
\text{Re}[\Gamma^-_{abcd}] = \frac{1}{\hbar}\sum_i \Lambda^i_{ab} \Lambda^i_{cd}  \begin{cases}
J(\omega_{ab}) \big [ n(\omega_{ab}) + 1 \big ] \ \ & \text{if} \ \ \omega_{ab} > 0 \\ 
J(\omega_{ba}) n(\omega_{ba})  \ \ & \text{if} \ \ \omega_{ba} > 0 \\ 
\lim\limits_{\omega \rightarrow 0} J(\omega) n(\omega) \ \ & \text{if} \ \ \omega_{ab} = 0 .
\end{cases}
\end{eqnarray}
For the Drude-Lorentz SD, we have $\lim\limits_{\omega \rightarrow 0} J(\omega) n(\omega) = 2 \lambda \frac{k_B T}{\hbar \omega_c}$. 
The real part determines the relaxation of the system.
The imaginary part can be expressed in terms of principal value integrals
and introduce terms that modify the transition frequencies. Such energy shifts are small and generally bring no qualitative contribution to the dynamics of $\sigma$,
therefore, the imaginary part is disregarded.

\section{Results and Discussion}

\subsection{System Preparation}

The parameters used in the simulations are presented in Table \ref{tbl:parameters}. 
Using these parameters, the set of coupled differential equations produced by Eq. \eqref{EQT1} are solved numerically by a fourth order adaptive step size Runge-Kutta method. 

\begin{table}
	\caption{Values and description of the parameters used in calculations}
	\label{tbl:parameters}
	\begin{tabular}{cccc}
\hline
		&     & value  & description \\
\hline
		$H^{\text{el}/\text{hl}}$        
		
		& $N$                & 84            & number of sites \\
		
		& $\ell_{\text{o}}$ & 0.5 nm        & confinement length for the empty site \\
		
		& $V_0$ & $-1.2$ eV & bare electronic coupling \\
		
		&  $d_{ij}$  & 1.3 nm   & distance of nearest neighbor sites \\
		
		& $\xi_{\text{bind}}$ & 0.2 eV   & electron-hole coupling constant \\
		
		$H_{SB}$ & $\lambda$ & 50 meV & reorganization energy \\
		
		& $\omega_c$  & 1334 $\text{cm}^{-1}$ & cutoff frequency \\
		
		$F^{res}$ & $\Omega$ &  272 $\text{cm}^{-1}$ & normal mode frequency \\
		&  $\mu$ & 12 u & site mass \\
		
		$F^{thermo}$ & $\gamma^{-1}$ & $0.1$ ps & phonon-phonon relaxation time \\
		& $T_B$        & 300 K         & bath temperature \\
\hline
		
	\end{tabular}
\end{table}

\begin{figure}[h!] 
\centering 
\includegraphics[scale=0.4]{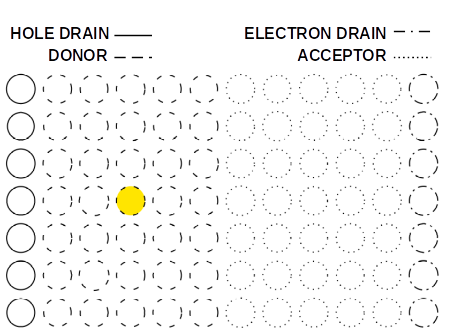}
\caption{Heterojunction comprised of 84 sites, consisting of a hole drain layer, donor material sites, acceptor material sites, and electron drain layer. 
The photoexcitation is considered to occur in the middle of the donor region, as depicted by the yellow circle in the figure. The radius of the molecular sites and the 
lattice parameter are drawn to scale.}
\label{init0}
\end{figure}

We consider a Frenkel exciton that has been photoexcited in the center of the donor region, as shown in Figure \ref{init0}. 
The figure uses the confinement length $\ell_i$ as the radius of the molecular sites. 
We consider a square lattice with lattice parameter $a$ = 1.3 nm. The effective electronic
coupling between nearest neighbors (nn) is set to be $V^{nn} = -40$ meV, as given by Eq. (\ref{hopping}), which is in agreement with other molecular crystal 
models.\cite{lee2015charge, coropceanu2007charge, mozafari2013polaron, kocherzhenko2015coherent, de2017vibronic, troisi2006dynamics,candiotto2017charge}
Likewise, the electronic coupling between next-nn sites is $V^{nnn}=-1.4$ meV, owing to the bigger distance. 
We also recall that the inter-molecular electronic couplings $V_{ij}$ are time-modulated by the form factor $F(\ell_{i},\ell_{j},d_{ij})$, due to the dynamics of the 
confinement radii $\ell$.
The coherent vibronic effects are produced by the coupling between the electronic degrees of freedom and $\ell$, via the Ehrenfest forces, 
as describes by Eq. (\ref{intra}). The dissipative vibronic effects are described by the Redfield relaxation tensor.
The fundamental vibrational mode of the C$_{60}$ molecule, Hg(1), of energy 272 cm$^{-1}$, is described classically because its energy satisfies the                    
condition $\hbar \omega/k_BT \approx$ 1.3, for T = 300K, and also due to its high symmetry. This mode, therefore, is responsible for the coherent (non-Markovian) 
vibronic effects in the model. In principle, other low energy molecular vibrational modes could be incorporated in the Ehrenfest dynamics, 
once the specific symmetry of the mode is taken into account. It is important to notice that this vibrational mode will also modulate the 
inter-molecular electronic coupling $V_{ij}$, since the intramolecular vibration changes the effective separation between molecular sites. 
The delocalized vibrational modes of the molecular crystal are assumed to be incoherent and weakly coupled to the frontier molecular orbitals. 
All the high energy coupling modes are treated quantum mechanically in the Redfield tensor, up to the cutoff frequency $\omega_c$ = 1334
cm$^{-1}$ that represents the localized C$=$C stretching mode.

Before performing the quantum dynamics simulations, we thermalize the system to obtain the initial configuration of the $(\ell_i, p_i)$  for 
each molecular site $i$.
To do so, we first run the dynamics with only the classical framework of the model, without the excitation of the electron-hole pair. 
Thus, the kinetic energy of the on-site confinement harmonic oscillator equilibrates with the bath temperature.

Once $\sigma^{\text{el/hl}}_{\text{adiabatic}}$ is obtained from equation Eq. \eqref{EQT1}, it is transformed to the diabatic local basis by 
$\sigma^{\text{el/hl}}_{\text{diabatic}} = \hat{U}^{\text{el/hl}} \sigma^{\text{el/hl}}_{\text{adiabatic}} (\hat{U}^{\text{el/hl}})^{-1}$, 
where $\hat{U}^{\text{el/hl}}$ is the unitary operator that diagonalizes $H^{\text{el/hl}}$. 
Although electrons and holes are represented by independent reduced density matrices, namely $\sigma^{\text{el}}$ and $\sigma^{\text{hl}}$, the dynamics of 
one affects the other via the electron-hole binding term of  Eq. \eqref{bind}. 

\subsection{Energetic Driving Force and Photovoltaic Efficiency}

The energy difference $E_g^{\text{opt}} - E_{CS} = E_{CB}(D) - E_{CB}(A)$ is considered the energetic driving force for charge generation in organic solar cells 
(OSC).\cite{Nakano2019,menke2018understanding} 
The energy $E_{CT}$ is alternatively used for this definition,\cite{menke2018understanding,Qian2018} but $E_{CT}$ in our model is given by the dynamics 
simulations, via $E_{\text{bind}}$, rather than being a parameter. Thus,  we use the former definition.
In this section we vary the energy offset $E_{DA}\equiv E_{CB}(D) - E_{CB}(A)$ to analyse its effect on the 
charge generation and on the photovoltaic conversion efficiency of an OSC.

\begin{figure}[h!]
	\centering
	\includegraphics[scale=0.56]{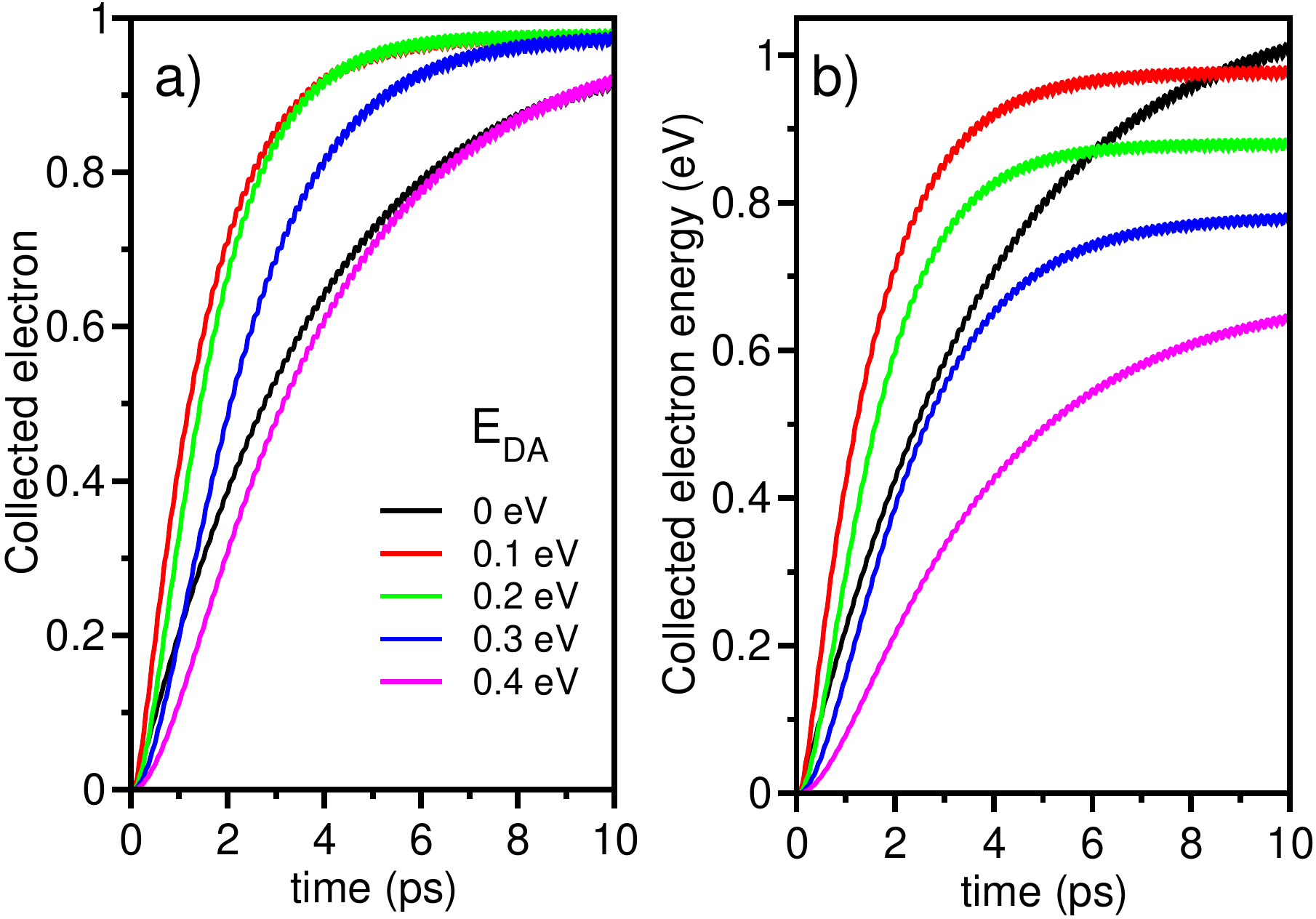}
	\caption{a) Electron collection in the drain layer for various  energy offsets $E_{DA}\equiv E_{CB}(D) - E_{CB}(A)$. 
b) Collected electron energy, here defined as the quantity $N_e \times E_{CS}$, where $E_{CS}=E_{CB}(A)-E_{VB}(D)$.}
	\label{gaps-comp}
\end{figure}

We varied $E_{DA}$ from 0 to 0.4 eV, while keeping the offset of the valence band fixed at 0.3 eV; notice that the photoexcitation is assumed to take place in 
the donor material and that recombination effects are not included in the model. The simulations were carried out with the quantum dissipative and Berendsen baths at 
$T_B$ = 300 K; no electric fields were applied.
Figure \ref{gaps-comp}-a) shows the electron population collected by the drain sites ($N_e$) as a function of $E_{DA}$.
According to the model simulations, the charge collection is optimal for 100 meV $\leq E_{DA} \leq$ 200 meV. 
Then, instead of increasing with the energetic driving force, the charge generation rate decreases for $E_{DA} \geq$ 300 meV. 
We ascribe this behavior to the Marcus inverted regime, whereby the electron-hole pair dissociation becomes unfavorable  as the Gibbs free energy of the process 
increases.\cite{Migliore2012,tamura2013ultrafast} 
Egorova et al. showed that the Redfield method is well suited to describe the Marcus inverted regime.\cite{egorova2003modeling}
The behavior has been observed for various combinations of donor/acceptor blends in planar heterojunctions.\cite{Atxabal2019,Nakano2019}
Additional insights on this effect are provided in the sections ahead, by analysing the temperature dependence of the charge generation process.
For $E_{DA} <$ 100 meV, on the other hand,  the electron generation rate is also small, in this case due to the lack of driving force.
By comparing the charge generation in fullerene and non-fullerene based solar cells, Qian et al.\cite{Qian2018} reported that the charge separation in 
polymer/fullerene blends occurs on a sub-300 fs timescale whereas for polymer/NFA blends it happens in timescales of 10-30 ps  for materials that 
exhibit $E_{DA} <$ 100 meV.
Next we consider the photovoltaic conversion efficiency associated with the different charge generation regimes

In Figure \ref{gaps-comp}-b) we estimate the ideal photovoltaic energy, $q\Delta V^{\text{ideal}}$, 
which we define as  $q\Delta V^{\text{ideal}}=N_e \times E_{CS}$, with $E_{CS}=E_{CB}(A)-E_{VB}(D)$.
It can be associated with the open-circuit voltage as $V_{OC} = \Delta V^{\text{ideal}} - \Delta V^{\text{losses}}$,
where $q \Delta V^{\text{losses}}$ represents the energy losses due to the various radiative and non-radiative recombination mechanisms. 
For the sake of argument, we also assume that $q \Delta V^{\text{losses}}$ is the same for all the cases considered. 
The exciton lifetimes vary from tens of picoseconds up to the nanosecond time scale, depending on the system characteristics,{\cite{Armin2021}} therefore, as a first approximation, we omit the recombination decay to describe the primary events of electron-hole separation dynamics.
Thus, for these simulations, the energetic driving force $E_{DA}$ = 100 meV provides the highest photovoltaic conversion as well as high charge generation rates 
during most of the simulation time.
Higher values of $E_{DA}$ render lower photovoltaic efficiencies due to higher energy losses at the interface. 
On the other hand, the case of $E_{DA}$ = 0, or $E_{DA} < $ 100 meV  for that matter, provides an interesting scenario.
For $E_{DA}$ = 0 the power conversion is small at short times, but it increases slowly until it reaches the highest collected energy for t $>$ 8 ps. 
This result shows that a small driving force can yield high power conversions in OSC, however, it also requires reduced recombination
rates since the charge generation rate is significantly lower for negligible $E_{DA}$. Indeed, it has been shown that NFAs used in high PCE solar cells exhibit  
long exciton lifetimes.\cite{classen2020role,Armin2021,Qian2018}

\subsection{Effect of vibronic Interactions on the Charge Separation Dynamics}

In this section we examine the influence of the vibronic couplings on the charge generation dynamics.
According to the electron-hole subsystem hamiltonian, Eq. (\ref{Hs}), the vibronic couplings have two origins in our model: the {\it intra}-molecular Holstein coupling,
governed by Eq. (\ref{intra}), and  the {\it inter}-molecular Peierls coupling produced by $V_{ij}$, defined in Eq. (\ref{hopping}). 
Both mechanisms are influenced by the classical thermal bath of temperature T$_B$ that acts on the vibrational coordinates. The {\it intra}-molecular Holstein 
interaction is also responsible for the coupling of the electron-hole subsystem with the quantum bath, through the Redfield relaxation tensor.

\begin{figure}[h!]
	\centering
	\includegraphics[scale=0.6]{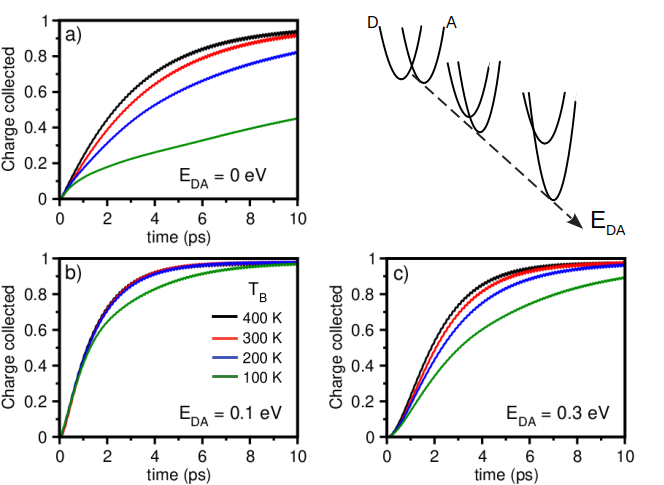}
	\caption{Electronic population ($N_e$) collected at the electron drain for various bath temperatures (T$_B$) and donor-acceptor energy offsets:
a) $E_{DA}$ = 0 eV, b) $E_{DA}$ = 0.1 eV and c) $E_{DA}$ = 0.3 eV.
The illustration in the top-right corner is a reference to the Marcus Theory, depicting electron transfer in the direct regime, the activationless regime, and the inverted regime as the 
$E_{DA}$ offset increases.} 
	\label{total-col-temperature} 
\end{figure}

The electronic population ($N_e$) collected at the electron drain is shown in Figure \ref{total-col-temperature}, 
for increasing energetic driving forces ($E_{DA}$) and various bath temperatures (T$_B$).
The model simulations show that the charge generation rate increases with temperature but,
furthermore, the effect of T$_B$ on $N_e$ reveals three distinct electron transfer regimes, as depicted in the inset of Figure \ref{total-col-temperature}:
a) the normal electron transfer for negligible $E_{DA}$ = 0 eV; b) the activationless regime for $E_{DA} \approx$ 0.1 eV; and c) the inverted Marcus regime for 
$E_{DA}$ = 0.3 eV. In general,  little difference is observed for T$_B$ = 300 K to 400 K, indicating that the charge generation rate saturates around room temperature.
In the activationless regime, for $E_{DA} \approx$ 0.1 eV, the process is almost independent of the temperature.
The biggest difference is observed for T$_B$ = 100 K ($\sim$ 8.6 meV), though, that shows 
significantly smaller rates for charge separation.
It is interesting to notice, in Figures {\ref{total-col-temperature}}-a) and c), that the charge transfer in the Marcus inverted region is higher than in the normal region, a behavior that is also revealed in the temperature dependence of the curves.   
This is in agreement with the Marcus-Jortner-Levich theory {\cite{Jortner,MJL,VSBatista}}, which predicts that vibrational mediated D-A electronic couplings 
decrease the effective size of the transition-state barrier. In our model, we ascribe this effect to the fact that the Redfield equation is cast onto the basis of 
the delocalized adiabatic eigenstates of $H_S$, thus mixing the diabatic on-site quantum states with the vibrational/bath modes.

The underlying processes responsible for the free charge generation in high-efficiency OSC are still a source of debate.\cite{Nakano2019}
Perdigón-Toro et at. identified a small activation barrier ($<$ 10 meV) for the charge generation in high performance PM6:Y6 non-fullerene 
blends.\cite{Andrienko}
Elsewhere, Gao et al. reported\cite{Gao2015}  activation energies of 9 meV for ordered P3HT:PC$_{60}$BM and of 25 meV for less ordered polymer:polymer 
interfaces.
On the other hand, Hinrichsen et al. suggest that, in NFA based solar cells with negligible driving forces, charge-transfer excitons  must overcome activation energy barriers of $\sim$ 200 meV to 
generate free charges over a time period of hundreds of picoseconds at room temperature.\cite{Hinrichsen2020}
Our simulations indicate that different tunings of the electron-transfer regime can lead to very different charge separation times, especially at low temperatures.

\begin{figure}[h!]
  	\centering
  	\includegraphics[scale=0.48]{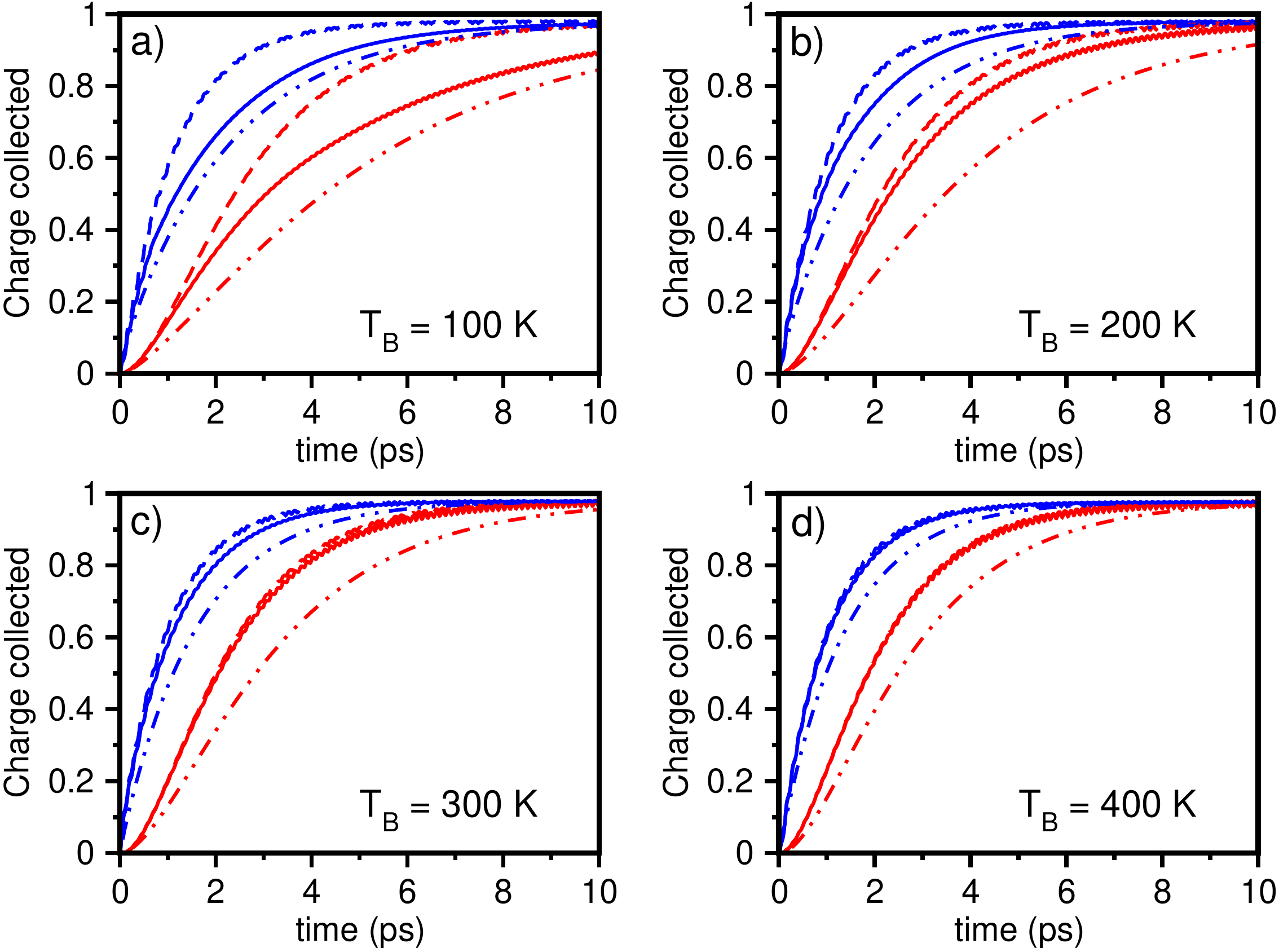}
  	\caption{Charge collected in the drain layers, $N_e$ (red) and $N_h$ (blue), at different bath temperatures (T$_\text{B}$):  
a) 100 K, b) 200 K, c) 300 K and d) 400 K. Donor-acceptor offset is $E_{DA}=$  0.3 eV.
{\bf Solid lines} describe charge dynamics considering both Holstein and Peierls mechanisms.
{\bf Dashed lines} describe charge collection with {\it inter}-molecular Peierls coupling, but without {\it intra}-molecular Holstein relaxation.
{\bf Dot-dashed lines} describe charge collection without vibrational effects.
} 
  	\label{col_charge} 
  \end{figure}

Although vibrational couplings tend to increase the charge generation yield, the different vibronic mechanisms play different roles.
Panel \ref{col_charge} shows the electronic, $N_e$ (red), and hole, $N_h$ (blue), densities collected at the respective drain layers. 
The hole densities are consistently higher because we assume that the photoexcited pair is created in the donor material, closer to the hole drain 
(see Figure \ref{init0}). Then, we consider three different scenarios for the vibronic coupling effects:
charge generation dynamics considering both the Holstein and Peierls mechanisms (solid-line curves); 
charge generation with {\it inter}-molecular Peierls coupling but disregarding the {\it intra}-molecular Holstein mechanism (dashed-line curves); 
charge generation without the Ehrenfest vibrational effects, that is, assuming rigid molecular sites with $\ell_i=\ell_0$ (dot-dashed curves). 
The Redfield relaxation tensor is taken into account for all of the above scenarios.
We also consider 
this situation for different bath temperatures T$_B$ and donor-acceptor offset of $E_{DA}=$ 0.3 eV
(see Appendix B for the cases $E_{DA}=$ 0 eV and 0.1 eV).

Now, we examine the different features revealed in the panels of Figure \ref{col_charge}. First, the lack of vibronic effects (dot-dashed curves) leads to less 
charge generation. This behavior is in accordance with the trend exhibited  in 
Figure \ref{total-col-temperature}, namely, the charge generation increases with the bath temperature, favoring endothermic processes. 
It corroborates studies\cite{bian2020vibronic, polkehn2018quantum, tamura2013ultrafast} that show that
the mixing of electronic and vibronic degrees of freedom improves the charge transport\cite{de2017vibronic} and, more importantly,
the vibrational effects help the electron-hole pair overcome the binding energy to give rise to the charge separated state.\cite{Nakano2019,Hinrichsen2020}
Interestingly, though, when we shut off the Holstein relaxation mechanism (dashed curve) the charge generation improves further in comparison with the full 
vibronic scenario (solid curve), particularly for the lower bath temperatures. 
We ascribe this effect to the polaron formation, that relaxes the carriers to a lower electronic energy state, rendering them less mobile. 
As the bath temperature T$_B$ increases, the carriers are forced to move due to the background energy of the bath and the Peierls mechanism.
Thus, for higher bath temperatures the Peierls mechanism overwhelms the polaronic effects, whereas for low T$_B$ the Holstein relaxation 
effect significantly 
hampers the charge generation. 
The enhancement of charge transfer due to the Peierls coupling mechanism, evinced in our simulations, is also reported in other studies.\cite{mozafari2013polaron, munn1985theory}
We recall that the effect is more pronounced for the electron, since it is created farther from the drain, on the opposite side of the interface.

We must take into consideration, however, that the mixed quantum-classical approach based on the Ehrenfest method 
tends to overestimate the occupation of electronic excited-states
in comparison with the surface hopping method,  for instance.\cite{parandekar2005mixed} 
Therefore, we have to keep in mind that the results shown in Figure \ref{col_charge} may overestimate the vibronic effects for the higher temperatures of the bath. 

\subsection{Electron-hole Binding}

\begin{figure}[h!]
    \centering
  \includegraphics[width = 7.0cm]{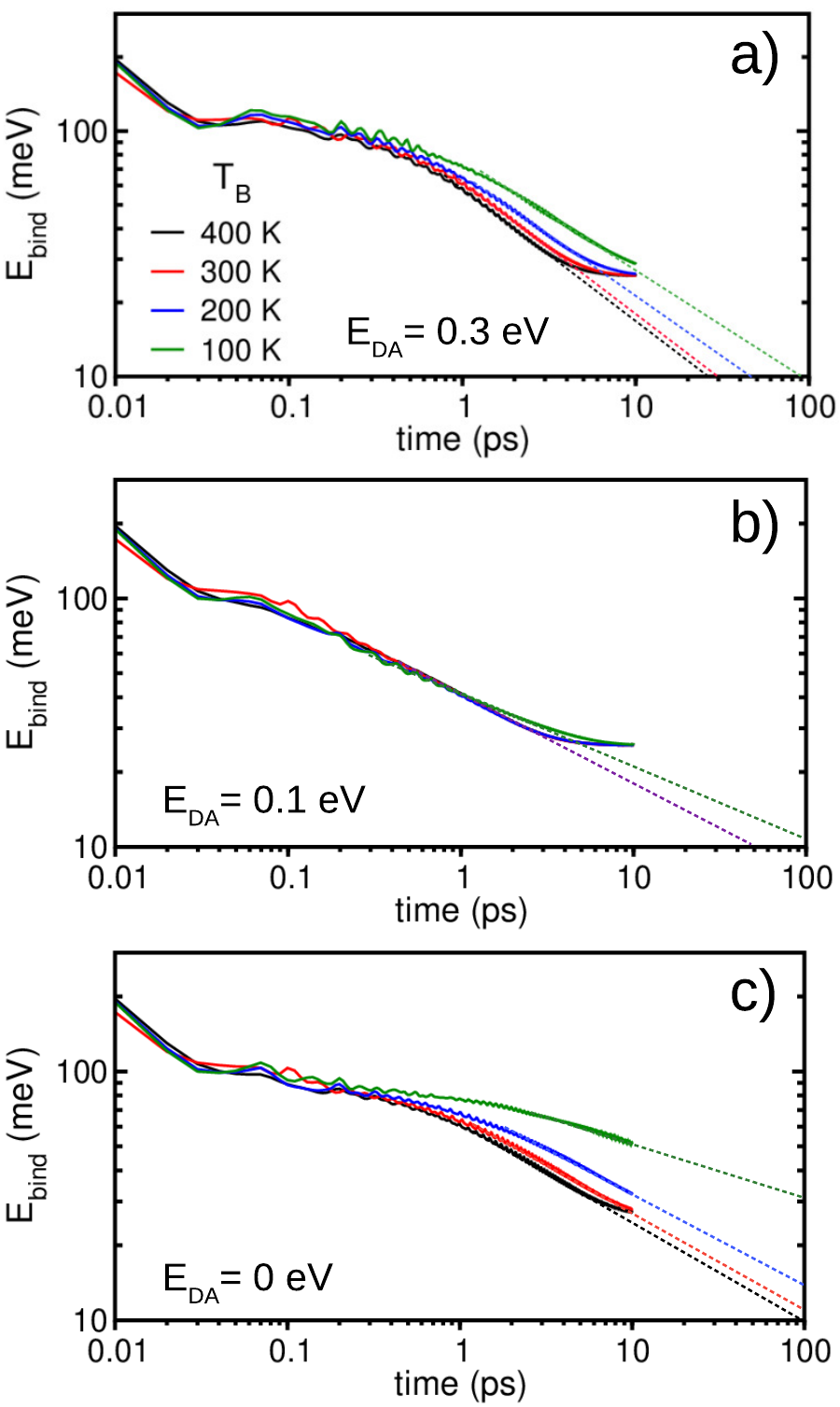}
  \caption{a) Electron-hole binding energy as function of time, as given by 
$E_{\text{bind}} = \text{Tr}[\sigma^\text{el} \Phi^\text{el}  + \sigma^\text{hl} \Phi^\text{hl}]$, for different D-A energy offsets:
a) $E_{DA}$ = 0.3 eV, b) $E_{DA}$ = 0.1 eV, and  c) $E_{DA}$ = 0 eV. The bath 
temperatures are T$_B$ = 100 K (green), 200 K (blue) , 300 K (red)  and 400 K (black).
The extrapolation of the curves is used to estimate the electron-hole separation time.} 
 \label{eh-bind} 
\end{figure}

We also analyse the electron-hole binding energy through the dynamics simulations, as evinced in 
Figure \ref{eh-bind} for different energetic driving forces ($E_{DA}$) and various bath temperatures (T$_B$). 
To do so, we examine the quantity $E_{\text{bind}} = \text{Tr}[\sigma^\text{el} \Phi^\text{el}  + \sigma^\text{hl} \Phi^\text{hl}]$, 
with $\Phi^\text{el/hl}$ given by Eq. (\ref{bind}). For the parameters given in Table \ref{tbl:parameters}, $E_{\text{bind}}(t=0) =$ 400 meV for all 
the cases considered, with the 
electron and hole occupying the same molecular site. For t = 10 fs,  $E_{\text{bind}}$ decreases to $\sim$ 200 meV, that corresponds to approximately 10\% of the electron 
and hole populations transferred to each of the 4 nearest neighbor molecular sites. As the dynamics proceeds, $E_{\text{bind}}$ decays by another half to 
approximately 100 fs, so that $E_{\text{bind}}(100 \text{ fs}) \approx $ 100 meV. 
The initial decay of $E_{\text{bind}}$ that occurs for $t\lesssim$ 100 fs is  practically independent of the bath temperature (T$_B$) and $E_{DA}$.
Thus, we ascribe this initial decay  to the excitonic spread in the donor material. Then, for $t > $ 200 fs, the decay of 
$E_{\text{bind}}$ becomes significantly slower and shows dependence on both T$_B$ and $E_{DA}$, indicating
the onset of the CT state dissociation.
In our simulations $E_{\text{bind}}$ saturates at approximately 25 meV, however, this is actually an artifact due to the finite size of the 
model. This value corresponds to the energy of the electron and hole populations trapped apart in their respective drain layers (see Figure \ref{init0}).

In Figure \ref{eh-bind} we also estimate the time the electron-hole pair would take to separate into free charges, by extrapolating the curves until 
$E_{\text{bind}}$ = 10 meV.
For $E_{DA}$ = 0.3 eV, it takes $\sim$ 30 ps for the binding energy between electron and hole to decrease to values much smaller than the available thermal energy at 
T$_B$ = 300 K. However, for the same energy offset and T$_B$ = 100 K, the time required for complete dissociation should exceed 100 ps.
For the activationless regime, Figure \ref{eh-bind}-b), the charge separation is practically temperature independent for T$_B >$ 200 K. 
Notice that the driving force and the temperature seem to assist the charge separation in Figure \ref{eh-bind}-a), though.
Finally, for negligible driving forces, Figure  \ref{eh-bind}-c), the final charge separation occurs much later, in addition to being strongly temperature dependent.
In particular, for T$_B <$ 200 K the charge separation should take several hundreds of picoseconds to occur, if at all.

It is important to notice that the specific behavior of the simulation is dependent on the model parameters, and the results 
reflect the parameters given in Table \ref{tbl:parameters}.
Even so, the simulations show that initially $E_{\text{bind}}$ looses most of its strength on a sub-100 fs timescale, so that when the dissociation of the CT 
state initiates $E_{\text{bind}}$
is actually much smaller than the exciton binding energy of the isolated molecular species. 

Generally, in order to simplify the description, the charge generation process is described as a succession of independent and sequential steps, such as: 
exciton diffusion to the interface, stabilization of the CT state, dissociation of the CT into CS state. 
However, the simulation results show that charge generation is facilitated if these mechanisms operate concomitantly.

\subsection{Effect of an Electric Field on the Charge Dissociation}

\begin{figure}[h!]
    \centering
  \includegraphics[scale=0.5]{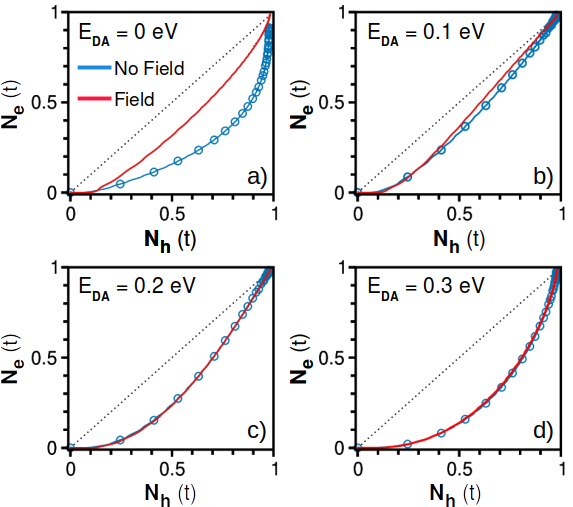}
  \caption{Parametric curves of $N_e(t)$ versus $N_h(t)$, with and without a macroscopic electric field at the interface region,
for various D-A energy offsets: a)  $E_{DA}$ = 0 eV, b)   $E_{DA}$ = 0.1 eV, c)   $E_{DA}$ = 0.2 eV and d)   $E_{DA}$ = 0.3 eV.
The diagonal dotted line indicates the asymptotic condition of instantaneous electron-hole pair dissociation.} 
 \label{Efield} 
\end{figure}

In the previous sections, we have focused on the intrinsic energetic properties of the D-A interface. Herein, we consider the influence of a macroscopic electric 
field $\mathcal{E}$ on the heterojunction. 
In general, the application of $\mathcal{E}$ induces two different effects: it contributes to the formation and subsequent dissociation of the CT-state 
-- by lowering the Coulomb potential barrier --
and it induces a drift of the free carriers toward the respective electrodes. Since both effects occur concomitantly, it is difficult to quantify them independently.
Experimentally, the electric field dependence of the CT-state photoluminescence can be used to determine the influence of $\mathcal{E}$ on the 
dissociation of the electron-hole pair \cite{Durrant2010,Nakano2019,Qian2018,Andrienko}. The drift effect, on the other hand, can be determined by measuring the 
mobility of the free carriers in the pristine materials\cite{coropceanu2007charge}. 
In any case, this is a controversial topic since various investigations report conflicting evidence over whether macroscopic electric fields commonly present in 
OSCs are sufficiently strong, or have any influence at all, on the charge photogeneration under operation 
conditions.\cite{Durrant2010,Nakano2019,Qian2018,Andrienko,Armin2021}

To obtain the electron-hole dynamics under the influence of an effective electric field on the D-A interface,
we include the additional term $q\mathcal{E}x$ to the energetic balance of the system hamiltonian, Eq. (\ref{Hs}), where 
$q=-e$ ($+e$) is the charge of the electron (hole) and $\mathcal{E}$ = 10 V/$\mu m$ is the amplitude of a constant electric field along the x direction.
It represents the net result of the built-in electric potential and, possibly, a reverse applied bias in the dielectric photoactive layer. 
The simulation results are shown in Figure \ref{Efield} by  parametric curves of the electron and hole populations in the respective drain layers, 
that is, $N_e(t)$ versus $N_h(t)$. We adopt this representation to distinct the effect of the electric field on the electron-hole pair dissociation
from the induced carrier drift.
First, we recall that the hole wavepacket drifts toward its drain layer without crossing the D-A interface (refer to Figure \ref{DAmodel}). 
Therefore, it is reasonable to assume that the main effect of $\mathcal{E}$ on $N_h(t)$ is due to drift. Moreover, since the photoexcitation takes place in the donor material,
the hole is created in the vicinity of its drain layer, thus rendering a short collection time.
Therefore, the parametric curves will always lie below the diagonal dotted line, which represents the asymptotic condition of instantaneous electron-hole pair 
dissociation. The more concave the parametric curve is, the longer it takes for the electron-hole pair to dissociate.
Then, considering the effect of $\mathcal{E}$ in the case of a vanishing $E_{DA}$ offset (Figure \ref{Efield}-a), 
we observe that the electric field strongly promotes the dissociation of the electron-hole pair.
However, if the intrinsic energetics of the system is already close to the activationless dissociation regime (Figure \ref{Efield}-b) the CT-state 
separation is already efficient 
and the electric field has a weak effect. 
Finally, for $E_{DA}$ = 0.2 and 0.3 eV, we have seen that the intrinsic energetics of the model puts the system in the inverted regime region for charge dissociation,
and the electric field reinforces the situation. 
In summary, in the case of small energetic driving forces a moderate electric field promotes the electron-hole pair dissociation by moving the energetics 
of the interface closer
to the crossover regime of the Marcus Theory. However, if the intrinsic energetic driving force already places the system close to the inverted regime, the electric
field should not improve the charge separation.

\section{Conclusions}

We have studied the influence of the energetic driving force and the vibronic effects on the charge generation of  
OSCs.  
The charge generation process is modelled by a coarse grained mixed quantum-classical model that provides relevant insights into the 
charge generation process in small energy offset interfaces. 
The simulations show that maximum charge generation occurs for a driving force $E_{DA}=E_g^{\text{opt}} - E_{CS} \approx$ 100 to 200 meV, 
corresponding to the activationless electron transfer regime.
The photovoltaic energy  conversion, on the other hand, is more sensitive to the energetic driving force and shows maximum 
efficiency for $E_{DA} \lesssim$ 100 meV. 
By analyzing the effects of the Holstein and Peierls vibrational couplings, we also examined the influence of vibronic couplings on the charge generation.
We find that vibrational couplings produce an overall effect of improving the charge generation. 
However, the two vibronic mechanisms play different roles. The Holstein relaxation mechanism decreases the charge generation
due to the formation the polaron, particularly for T$_B < $ 200 K. 
The Peierls mechanism always assists the charge generation and, for T$_B > $ 200 K,
it tends to overwhelm the Holstein mechanism effects.
Since the model does not take into account the recombination losses, the influence of this mechanism is considered only qualitatively. 
Finally, by examining the time-dependent electron-hole binding energy, we evince two distinct regimes for charge dissociation: the temperature independent 
excitonic spread on the sub-100 fs timescale and the CT state dissociation on the timescale of tens of picoseconds, so that when the electron-hole
pair reaches the interface its binding energy is much smaller that the initial excitonic binding energy.
In the presence of a macroscopic electric field, we find for
the case of small D-A energy offsets that a moderate electric field promotes the electron-hole pair dissociation by moving the energetics 
of the interface closer to the activationless regime. However, if the energy offset at the interface is large the macroscopic electric field does not 
improve the charge separation.

\begin{acknowledgements}
AMA is grateful for the financial support from FAPESC (Fundação de Amparo à Pesquisa do Estado de Santa Catarina). 
LGCR acknowledges support  by Coordena\c c\~ao de Aperfei\c coamento de Pessoal de N\'{\i}vel Superior Brasil (CAPES) - Finance Code 001, 
by the Brazilian National Counsel of Technological and Scientific Development (CNPq) and the National Institute for Organic Electronics (INEO).
The authors acknowledge support by allocation of supercomputer time from Laboratory for Scientific Computing (LNCC/MCTI, Brazil).
\end{acknowledgements}

\section*{Data Availability Statement}
The data that supports the findings of this study are available within the article.

\appendix

\section{Ehrenfest Force}

Writing the time dependent wave function for the electron and hole in terms of the diabatic basis of the system, we have
\begin{equation}
| \Psi^{\text{el/hl}}(t) \rangle = \sum_i C^\text{el/hl}_{i} (t)| i \rangle 
\end{equation}
where the elements of reduced density matrix are defined $\sigma^{\text{el/hl}}_{ij} =  C^{\text{el/hl}}_i (C^{\text{el/hl}}_j)^* $. In the next equations, the superscripts for electron and hole will be omitted for the sake of clarity. The RHS of Eq. \eqref{Ehrenfest1} turns

\begin{equation}
\frac{\partial U}{\partial \ell_k} = \sum_{i, j} \Big \{ H_{ij}  \Big (  \frac{\partial C^*_i(t)}{\partial \ell_k} C_j(t) + C^*_i(t)  \frac{\partial C_j(t)}{\partial \ell_k} \Big ) + C^*_i(t) \frac{\partial H_{ij}}{\partial \ell_k} C_j(t) \Big \} 
\end{equation}

Using the time dependent Schrödinger equation, we have

\begin{equation}
\frac{d C_{j}(t)}{d\ell_k} = - \frac{i}{\hbar \dot{\ell_k}} \sum_{n} C_{n}(t) H_{j n} - \sum_{n} C_{n}(t) \langle {j | \frac{\partial n}{\partial \ell_k}} \rangle 
\end{equation}

Thus 
\begin{eqnarray}
\sum_{i, j} \Big \{ H_{ij}  \Big (  \frac{\partial C^*_i(t)}{\partial \ell_k} C_j(t) + C^*_i(t)  \frac{\partial C_j(t)}{\partial \ell_k} \Big ) = \frac{i}{\hbar \dot{\ell}_k} \sum_{i, j, n} H_{ij} (H_{in} \sigma_{jn} - H_{jn} \sigma_{ni} ) - \nonumber \\
- \sum_{i, j, n} H_{ij} \Big (\sigma_{jn} \langle i | \frac{\partial n}{\partial \ell _k} \rangle  + \sigma_{ni} \langle j | \frac{\partial n}{\partial \ell_k} \rangle         \Big ) 
\label{first_term}
\end{eqnarray}
The term $\sum_{i, j, n} H_{ij} (H_{in} \sigma_{jn} - H_{jn} \sigma_{ni} )$ is zero, what can be proven using the cyclic property of trace. We also suppress the second term in RHS of equation \eqref{first_term}, since we are considering the non-orthogonality only with the form factor $F(\ell_i, \ell_j, d_{ij})$ term. Thus
\begin{equation}
\frac{\partial U}{\partial \ell_k} = \sum_{i, j} C^*_i(t) \frac{\partial H_{ij}}{\partial \ell_k} C_j(t) = \sum_{i = j} \sigma_{ii} \frac{\partial \epsilon_i }{\partial \ell_k } + \sum_{i \neq j} \sigma_{ij} \frac{\partial V_{ij}}{\partial \ell_k}~.
\end{equation}

\section{Vibronic Interactions: Normal and Activationless Marcus Regimes}

\begin{figure}[h!]
    \centering
  \includegraphics[scale=0.3]{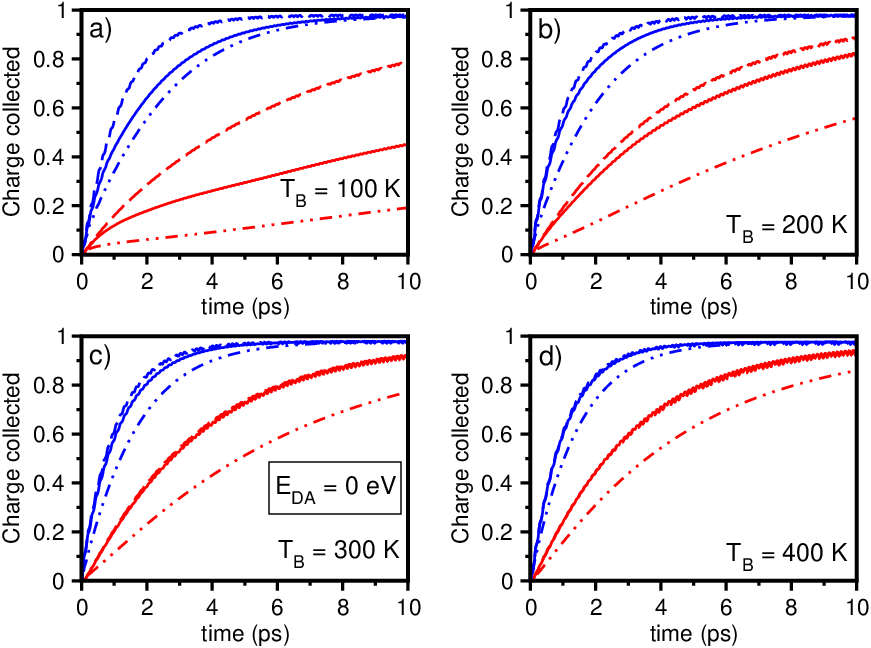}
  \caption{Charge collected in the drain layers: $N_e$ (red) and $N_h$ (blue).  
The bath temperatures (T$_\text{B}$) are:  a) 100 K, b) 200 K, c) 300 K and d) 400 K. The donor-acceptor offset is $E_{DA}=$  0 eV.
{\bf Solid lines} describe charge dynamics considering both Holstein and Peierls mechanisms.
{\bf Dashed lines} describe charge collection with {\it inter}-molecular Peierls coupling, but without {\it intra}-molecular Holstein relaxation.
{\bf Dot-dashed lines} describe charge collection without vibrational effects.
} 
 \label{F9} 
\end{figure}

\begin{figure}[h!]
    \centering
  \includegraphics[scale=0.3]{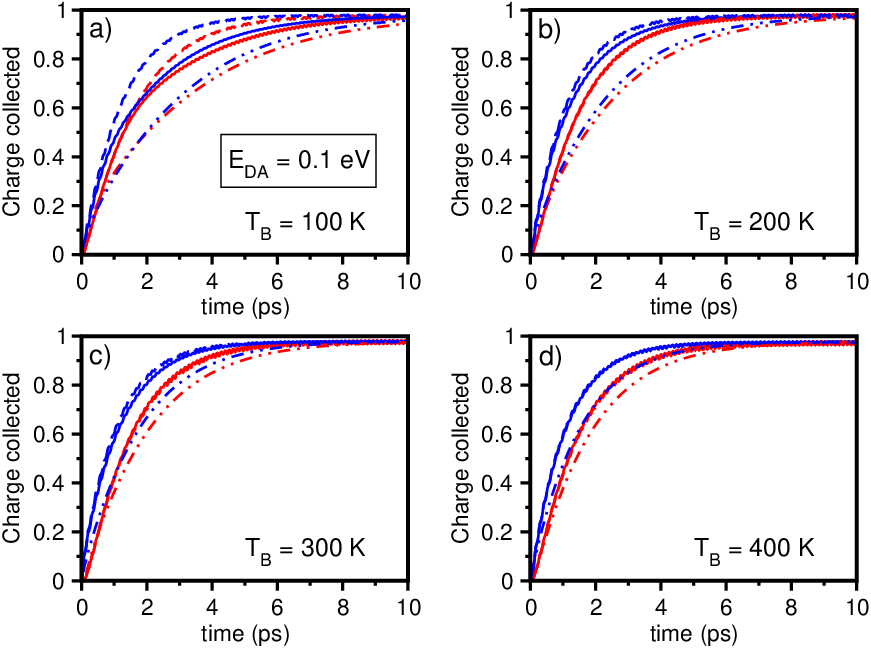}
  \caption{Charge collected in the drain layers: $N_e$ (red) and $N_h$ (blue).  
The bath temperatures (T$_\text{B}$) are:  a) 100 K, b) 200 K, c) 300 K and d) 400 K. The donor-acceptor offset is $E_{DA}=$  0.1 eV.
{\bf Solid lines} describe charge dynamics considering both Holstein and Peierls mechanisms.
{\bf Dashed lines} describe charge collection with {\it inter}-molecular Peierls coupling, but without {\it intra}-molecular Holstein relaxation.
{\bf Dot-dashed lines} describe charge collection without vibrational effects.
} 
 \label{F10} 
\end{figure}

The following panels show the effect of vibronic interactions on the charge separation dynamics for a vanishing $E_{DA}$ offset  (Fig. \ref{F9}) and for 
$E_{DA}$ = 0.1 eV (Fig. \ref{F10}). 
We associate the dynamics that occurs on these energetic conditions with the normal and activationless regimes of the Marcus electron transfer kinetics, respectively.
The condition associated with the inverted regime ($E_{DA}$ = 0.3 eV) is discussed in the main text of the paper. 

\clearpage

\bibliography{MyRefs}

\end{document}